\documentclass{aa}

\usepackage{graphicx}
\usepackage[varg]{txfonts}
\usepackage{pdfpages}
\usepackage{soul}
\usepackage{ulem}
\usepackage{xcolor}

\begin{document}

\title{The low-frequency break observed in the slow solar wind magnetic spectra}

\titlerunning{Slow wind turbulence spectrum}
\authorrunning{Bruno et al.}

%\correspondingauthor{Roberto Bruno}
%\email{roberto.bruno@inaf.it}

\author{R. Bruno\inst{1} \and D. Telloni\inst{2} \and L. Sorriso-Valvo\inst{3,}\inst{4} \and R. Marino\inst{5} \and R. De Marco\inst{1} \and R. D'Amicis\inst{1}}

\institute{National Institute for Astrophysics, Institute for Space Astrophysics and Planetology, Via del Fosso del Cavaliere 100, 00133 Roma, Italy
\and
National Institute for Astrophysics, Astrophysical Observatory of Torino, Via Osservatorio 20, 10025 Pino Torinese, Italy
\and
Departamento de F\'isica, Escuela Polit\'ecnica Nacional, Quito, Ecuador
\and
Nanotec/Consiglio Nazionale delle Ricerche, ponte P. Bucci, cubo 31C, I-87036 Rende (CS), Italy
\and
Laboratoire de M\'ecanique des Fluides et d'Acoustique, CNRS, \'Ecole Centrale de Lyon,
 Universit\'e Claude Bernard Lyon~1, INSA de Lyon, F-69134 \'Ecully, France
}

\date{Received ; accepted }

\abstract{Fluctuations of solar wind magnetic field and plasma parameters exhibit a typical turbulence power spectrum with a spectral index ranging between $\sim 5/3$ and $\sim 3/2$. In particular, at $1$ AU,  the magnetic field spectrum, observed within fast corotating streams, also shows  a clear steepening for frequencies higher than the typical proton scales, of the order of $\sim 3\times10^{-1}$ Hz,  and a flattening towards $1/f$  at frequencies lower than  $\sim 10^{-3}$ Hz.  However, the current literature reports observations of the low-frequency break only for fast streams. Slow streams, as observed to date, have not shown a clear break, and this  has  commonly been attributed to  slow wind intervals  not being long enough.
%\LEt{ ok like so? (I had to rearrange to avoid a syntax problem) }
Actually, because of the longer transit time from the Sun, slow wind turbulence would be older and the frequency break would be shifted to lower frequencies with respect to   fast wind. Based on this hypothesis, we performed a careful search for long-lasting slow wind intervals throughout $12$ years of Wind %\sout{WIND}
satellite measurements.
%\LEt{ in the acknowledgements and figures you use WIND. }
Our search, based on stringent requirements not only on wind speed but also on the level of magnetic compressibility and Alfv\'{e}nicity of the turbulent fluctuations,  yielded  $48$ slow wind streams lasting longer than $7$ days. This result allowed us to extend our study to frequencies sufficiently low and, for the first time in the literature, we are able to show that the $1/f$ magnetic spectral scaling is also present in the slow solar wind, provided the interval is long enough. However, this is not the case for the slow wind velocity spectrum, which keeps the typical Kolmogorov scaling throughout the analysed frequency range.
After ruling out the possible role of compressibility and Alfv\'enicity for the $1/f$  scaling, a possible explanation in terms of magnetic amplitude saturation, as recently proposed in the literature, is suggested.}


\maketitle
%
%
%%%%%%%%%%%%%%%%%%%%%%%%%%%%%%%%%%%%%%%%%%%%%
\section{Introduction}
\label{sec:intro}
%%%%%%%%%%%%%%%%%%%%%%%%%%%%%%%%%%%%%%%%%%%%%
%
The properties of heliospheric magnetic field and particle velocity fluctuations have been studied for decades with the major aim of understanding the mechanisms that govern the dynamics of the collisionless solar system plasmas~\citep{brunocarbone2013}.
The measurements provided by several spacecraft have allowed us to determine that the solar wind expansion is highly turbulent, as suggested by the ubiquitous observation of Kolmogorov-like magnetic field and velocity power spectra~\citep{K41,frisch,tu1995,brunocarbone2013,tsurutani2018}.

At the low-frequency end of the spectra, the fast and Alfv\'{e}nic solar wind
often displays a robust $1/f$ scaling range ($f$ being the frequency), which has been interpreted as injection range for the turbulent energy cascade. This was observed both in the ecliptic and in the polar solar wind~\citep{bruno2009, horbury1996, matthaeus2007}, for both magnetic field and velocity, but the reason for the formation of such scaling is still an open question.

Indeed, a range over which the low-frequency spectrum
%\LEt{ I was not sure here. Is this (low-frequency) spectrum [a spectrum that is low frequency] =or= low (frequency spectrum) [a frequency spectrum that is low]? in the end I put in the hyphen, but if it is the latter, please take it out, but only for this expresion.     }
of physical quantities follows power-law scalings close to $-1$ is observed in a variety of turbulent systems, for instance in geophysical fluids \citep{Fraedrich, Costa}, in  flow simulations in laboratory \citep{Herault,Pereira}, and in numerical simulations of hydrodynamic (HD) and magnetohydrodynamic (MHD) flows \citep{dmitruk2007}.

Moreover, evidence of long-period fluctuations associated with the $1/f$ low-frequency spectrum have been provided by \citet{dmitruk2011} in HD and MHD systems allowing condensation of invariants (or quasi-invariants) at the lowest wave-number mode. This is the case for three-dimensional MHD plasmas (with or without a background magnetic field) and for rotating HD flows where inverse cascades of helicity or energy can indeed develop and have been associated with the onset of a $1/f$ low-frequency spectrum in direct numerical simulations \citep{dmitruk2011}.

However, limiting ourselves to interplanetary space plasmas, it is worth  mentioning some past and current views on the nature of the $1/f$ scaling range.
%\LEt{ Single-sentence paragraphs should not be used and should be included in the previous or following paragraph, as appropriate. Please check throughout. }
For example, \citet{nakagawa1974} pointed out the possible link between the $1/f$ scaling in the interplanetary magnetic field and the structured surface of the Sun, based on early observations of a clear $1/k$ ($k$ being the wavenumber)
%\sout{wavelength})
spectral region in the solar photospheric magnetograms.
Instead, \citet{matthaeus1986} proposed that this kind of scaling could be the result of the early (i.e. within the Alfv\'enic radius) superposition of uncorrelated samples of turbulence, whose correlation lengths are lognormally distributed, each originating at a different regions of the solar surface. As shown by \citet{montroll1982}, a similar superposition of turbulence samples, in certain circumstances, can produce a $1/f$ scaling.
A different interpretation was given in terms of outward travelling
%\LEt{ I have corrected for UK language conventions in this paper. }
low-frequency waves propagating from the coronal base, whose superposition could generate the $1/f$ magnetic field and density spectrum. In this framework, the $1/f$ range was not present in purely hydrodynamic simulations, suggesting the central role of the magnetic field in its production~\citep{dmitruk2002, dmitruk2004}.
\citet{consolini2015}, using numerical simulations based on a shell model, obtained the formation of such a spectral domain in both fluid and MHD cases. These authors explained the emergence of such a domain in terms of a competition between direct and inverse energy cascading at the sub-inertial scales.
Recent studies point to the possibility that the emergence of coherent structures and/or the condensation of energy on large scales could be  the origin of the  $1/f$ low-frequency spectrum in fluid and plasma turbulence. In particular, it has been shown in liquid metal experiments (at high Reynolds number) that  three-dimensional shear flows and quasi-two-dimensional flows both exhibit a low-frequency $1/f$ spectrum, due to an increase in power in the gravest modes, caused respectively by the instability of the shear layer \citep{Pereira} and the onset of a large-scale circulation \citep{Herault}.

\citet{velli1989} and then \citet{verdini2012} and \citet{tenerani2017} suggested that outward propagating modes could be reflected by large-scale solar wind gradients in the extended solar corona, and their non-linear interaction would result in a turbulent cascade with spectral scaling $1/f$, already visible within the sub-Alfv\'enic solar wind~.
The latest interpretation is due to \citet{matteini2018} who suggested that the $1/f$ spectrum could be due to the saturation of the magnetic field Alfv\'enic fluctuations to the limiting value represented by the magnetic magnitude.

The analysis of the magnetic field spectral properties in the expanding solar wind showed a clear radial evolution of the frequency of the break between the $1/f$ region and the typical $\sim f^{-5/3}$ Kolmogorov fully developed turbulence. In particular, the break frequency decreases with the heliocentric distance $R$ roughly as a power law $\sim R^{-1.52}$ for fast ecliptic wind \citep{brunocarbone2013} ($\sim R^{-1.1}$ for fast polar wind \citep{horbury1996}).
A similar evolution, but with a slower decrease ($\sim R^{-1.1}$), was also observed for the spectral break between the Kolmogorov range and the high-frequency kinetic range~\citep{bruno2014a}. These observations suggested that the turbulence could develop as the solar wind travels away from the Sun, involving progressively larger scales into the turbulent cascade. This radial evolution increases the extension of the inertial range of solar wind turbulence, and consequently of the Reynolds number~\citep{matthaeus2005, telloni2015}.

Contrary to the fast solar wind, the analysis of the spectra of  slow wind
%\sout{winds}
(i.e. solar wind streams with bulk speed $\lesssim 400$ km s$^{-1}$)  provided evidence of Kolmogorov scaling all the way down to the low-frequency range~\citep{brunocarbone2013}. Based on our present interpretation of the observed phenomenology and on the basis of available models, this is not surprising since the slow wind turbulence has more time to develop during its slower expansion, thus allowing the possible low-frequency spectral break to drift towards lower frequencies than in the fast wind.
If this is the case, then we can expect to be able to observe the low-frequency break only in particularly long wind samples, where the low-frequency spectral properties can be properly captured.

In this paper we provide the first evidence of the existence of the low-frequency break and of a $1/f$ spectral scaling in a selection of slow wind samples. We then discuss the possible origin of this scaling in the framework of the interpretations listed above.
%
%%%%%%%%%%%%%%%%%%%%%%%%%%%%%%%%%%%%%%%%%%%%%
\section{Data analysis}
\label{sec:data}
%%%%%%%%%%%%%%%%%%%%%%%%%%%%%%%%%%%%%%%%%%%%%
%
In order to study the low-frequency spectral properties of the slow solar wind, we performed a systematic search for slow wind intervals using $12$ years' worth of solar wind observations recorded by the Wind spacecraft
%\LEt{ see note 2 }
between $2005$ and $2016$.
The following plasma and magnetic field data sets were used throughout the analysis: the Wind 3DP, PESA-LOW \citep{lin1995} onboard computed ion moments (proton and $\alpha$ particles) $3$ sec (spin) resolution, and the Wind Magnetic Field Investigation (MFI) experiment \citep{lepping1995} $3$ sec averages, respectively.
Considering the slight changes in the plasma sampling time during the mission, these two data sets were interpolated and re-sampled with a six-second cadence in order to allow for a synchronized study.

%$-----------------------------------$

%\LEt{ see note 4 }
Our selection of slow wind intervals was based on the evaluation of the following parameters: wind speed, time duration, magnetic intensity variability and Alfv\'{e}nicity.
The solar wind speed had to be consistently small ($V_{\rm{SW}}\,\lesssim\,400$ km s$^{-1}$) for an interval of at least 7 days  in order to extend our study to  sufficiently low frequencies. The time interval should not contain strong transient events or shocks which would alter the magnetic field compressibility
and the average value of the normalized standard deviation of the field intensity $\sigma_B/|B|$, estimated at hourly scale, should not exceed the value of $0.15$.
Finally, the Alfv\'{e}nicity of the fluctuations, estimated at hourly scales by the normalized cross-helicity $\sigma_{c}$ (defined as $2\,\delta \mathbf{V}\cdot \delta \mathbf{B}/(E_{v}+E_{B})$, where $\delta \mathbf{V}$ and $\delta \mathbf{B}$ are the velocity and magnetic field vector fluctuations and $E_{v}$ and $E_{b}$ are the kinetic and magnetic energy, respectively) should be less than $0.5$. This last requirement allows us to exclude from our data set the Alfv\'{e}nic slow wind, recently studied by \cite{raffaella2015} and \cite{raffaella2019}, which is more similar to the fast Alfv\'{e}nic wind than to the slow wind analysed in this work.

These selection criteria allowed us to identify 48 time intervals, which we found to be randomly distributed within the 12 years under study.
The resulting data set was primarily used to evaluate solar wind velocity spectra and the Alfv\'{e}nicity of the fluctuations within each of the selected time intervals. Additionally, given the better resolution and completeness of the magnetic field measurements, magnetic spectra, and magnetic compressibility were evaluated using the three-second cadence data, providing a more robust estimate.

\begin{figure*}
        \centering
        \includegraphics[width=15cm]{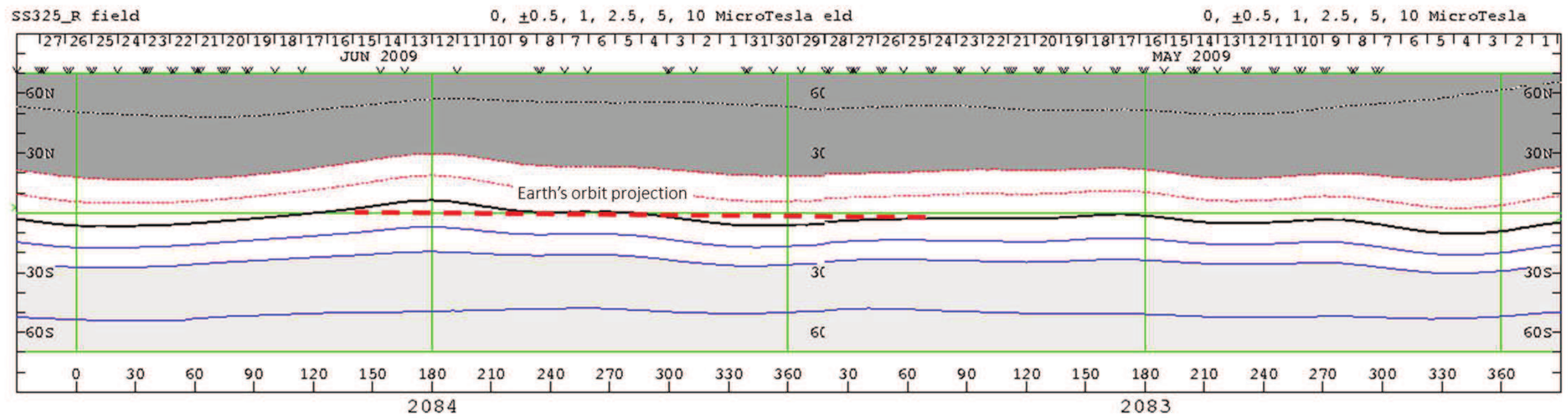}
        \caption{Source surface synoptic maps of Carrington Rotation 2083 and 2084 from Wilcox Solar Observatory as inferred at $3.25$ solar radii. Light blue shading shows the positive regions. The neutral line is black. The dashed red line represents the Earth's orbit back-projected onto the Sun using daily values of solar wind speed.}
        \label{fig:f1}
\end{figure*}

Among the 48 selected cases, we show one example relative to one of the longest and most representative streams, recorded from day $150$ to $173$ of $2009$. During this $23$-day interval, the Earth's heliographic latitude changed from $-1.5^{\circ}$ to $+1.14^{\circ}$. Because of the specific configuration of the heliomagnetic equator during the observation time, the Earth was steadily close to the ecliptic current sheath during the whole time interval.
This particular configuration can be observed in Fig.~\ref{fig:f1}, showing the source surface synoptic maps of Carrington Rotation $2083$ and $2084$ from the Wilcox Solar Observatory, as inferred at $3.25$ solar radii.
%\sout{The  light blue shading shows the positive magnetic field polarity regions. The neutral line is marked in black, while the dashed red line represents the Earth's orbit back-projected onto the Sun using daily values of the solar wind speed.}
%\LEt{ Legend information (colors, lines, etc.) should not be repeated in the main text unless necessary to understand the discussion.  }

Some relevant solar wind parameters relative to this time interval are shown in the shaded area of panel a of Fig.~\ref{fig:f2}, which  spans from day $120$ to $200$ of year $2009$.
Such extremely long time intervals, characterized by
%\sout{a}
an almost steady slow wind speed, are uncommon and this explains the reason for the limited number of intervals that satisfy our search criteria within 12 years of data.
The speed profile of the selected interval (shaded area) does not show large variability, with an average value of $316\pm 20$ km s$^{-1}$, while the other parameters show the typical variability of slow wind.
Panel b of Fig.~\ref{fig:f2} shows the total power density spectrum, i.e. the trace of the spectral matrix, normalized to the square value of the mean magnetic field intensity. This figure represents
the first evidence of the existence of the low-frequency break within the slow solar wind. The vertical dashed line separates the high-frequency range, characterized by about three decades of typical quasi-Kolmogorov scaling with exponent close to -5/3, and the low-frequency range, showing a power-law scaling with exponent close to -1 and extending for more than two decades.
To the best of our knowledge, this is a new observation that was never reported in the literature.
In the example shown here, the low-frequency spectral break is located around $10^{-4}$ Hz, about one order of magnitude lower
%\LEt{ hertz is a rate, yes? related to frequency? low/high }
than the typical values observed in the fast solar wind, closer to $10^{-3}$ Hz~\citep{brunocarbone2013}. This implies that particularly extended intervals are necessary in order to observe the $1/f$ scaling in the slow solar wind.

It is interesting to compare this value with the break location predicted by the radial dependence $R^{-1.52}$ valid for the fast wind in the ecliptic \citep{brunocarbone2013}. To do this, we analyse a typical fast wind interval highlighted by the dashed area in panel a of  Fig. \ref{fig:f3}. This is a typical corotating, high-velocity stream characterized by an average speed of $642\pm 44$ km s$^{-1}$.
Panel b  of Fig. \ref{fig:f3} shows the relative trace of the power density spectral matrix of the magnetic field fluctuations. This spectrum, as expected, shows a frequency break around $10^{-3}$ Hz, a typical value for fast wind \citep{brunocarbone2013}.
A lower expansion speed implies a longer transport time which, in turn, implies older turbulence. Given the linear relationship between transport time, velocity, and radial distance we can estimate the frequency break location at $1$ AU for our slow wind from the frequency break of the fast wind and the radial dependence reported by \cite{brunocarbone2013}:

\begin{equation}
f_{\rm{slow}}=f_{\rm{fast}}(V_{\rm{fast}}/V_{\rm{slow}})^{-1.52}=3.4 \times 10^{-4}\textrm{Hz}
\label{eq_radial}\end{equation}

The estimate from Eq. (\ref{eq_radial}) provides a value higher than $10^{-4}$ Hz shown in Fig. \ref{fig:f2}. Although this value should be taken  as a rough estimate, we verified that this discrepancy is not an isolated case, but generally applies to the break location observed for the slow wind at $1$ AU, although within a certain variability. Thus, the location of the slow wind break does not seem to be regulated by the age of turbulence as estimated from Eq. (\ref{eq_radial}), and the reason for this discrepancy might have a different origin.

However, a long enough time interval of slow wind seems to be a necessary but not sufficient condition  to have a clear low frequency break. In  panel a of Fig.~\ref{fig:f4}, we show another example of slow wind interval where the low-frequency break is clearly absent, in spite of the remarkably long duration of this sample.
The example shown here refers to a slow wind time interval lasting seven days and moving with an average flow speed of $305\pm 35$ km s$^{-1}$. The magnetic field power spectrum is shown in panel b of the same figure, and is characterized by a typical Kolmogorov scaling throughout the whole frequency range, for about four decades, with no observable low-frequency break up to scales of about one day. In this case, Eq. (\ref{eq_radial}) would predict a frequency break at $3.2~\times 10^{-4}$ Hz.

A remarkable difference between the two examples of slow wind shown in Figs. \ref{fig:f2} and \ref{fig:f4} is found in the normalized spectral power level. Indeed, the spectrum without a low-frequency break (Fig.~\ref{fig:f4}) has much less power than the one with the break (Fig.~\ref{fig:f2}). This observation is particularly relevant and is discussed in the following paragraphs.

We  compared  other turbulence aspects of these two time intervals. In particular, we looked at magnetic compressibility and Alfv\'{e}nicity. To evaluate the compressibility we estimated the ratio between the power associated to magnetic field intensity fluctuations and the total magnetic energy, namely the trace of the spectral matrix, $C(f)=E_{|B|}(f)/\sum_{i=x,y,z} E_{b_i}(f)$~\citep{bavassano1982}.
On the other hand, the degree of Alfv\'{e}nicity was evaluated computing, as is customary,  the normalized cross-helicity $\sigma_{c}(f)=(e^{+}(f)-e^{-}(f))/(e^{+}(f)+e^{-}(f))$ in terms of Els$\ddot{\textrm{a}}$sser variables, where $e^{+}(f)$ and $e^{-}(f)$ are the total power associated with outward and inward Alfv\'{e}nic fluctuations, respectively~\citep{brunocarbone2013}. In Fig.~\ref{fig:f5} we show plots of $C(f)$ and $\sigma_{c}(f)$ versus frequency, for the two time intervals shown in Figs.~\ref{fig:f4} and~\ref{fig:f2}. These two parameters do not display substantial differences for the two time intervals: the level of magnetic compressibility is low and very similar (panels a and b), and the Alfv\'{e}nicity (panels c and d) shows only a rather modest difference, being slightly larger
%\LEt{ higher if a rate, larger if a size }
for the second interval not characterized by the frequency break. Such a difference, however, does not seem to explain the absence of the spectral break for the corresponding time interval, since enhanced Alfv\'{e}nicity and low compressibility, as shown in panels e and f relative to the fast stream of Fig.~\ref{fig:f3}, are always associated with the clear presence of a low-frequency break~\citep{brunocarbone2013}.

\begin{figure}
\includegraphics[width=8cm]{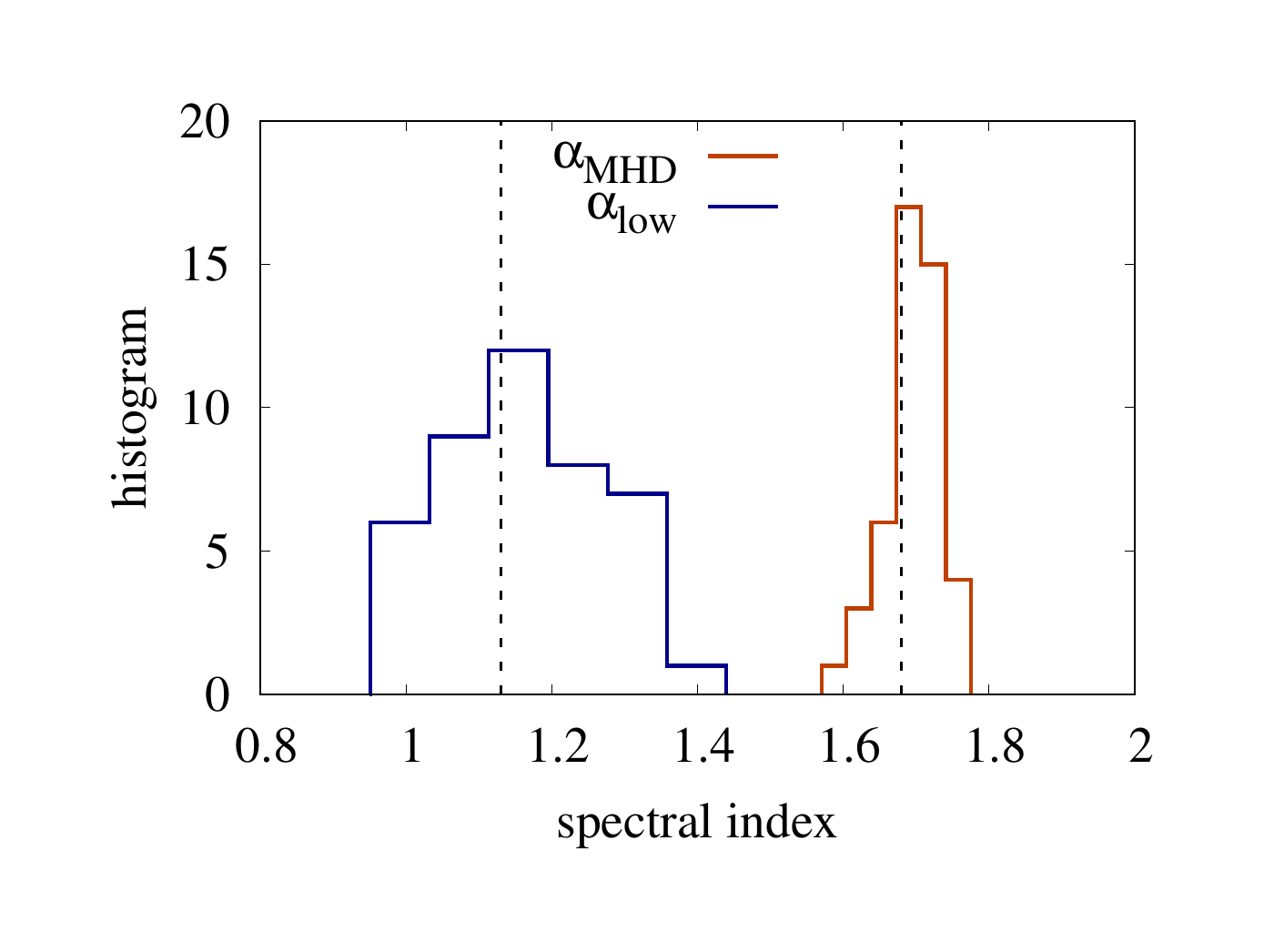}
\caption{Histograms of the power spectral exponents in the low-frequency (blue) and inertial (dark orange) ranges, as obtained for the 48 slow solar wind intervals measured by Wind
%\sout{WIND}
and selected for this work. The vertical dashed lines indicate the average within each frequency range.} \label{fig:f6}
\end{figure}

The analysis of the magnetic spectral properties of the $48$ selected intervals robustly shows the presence of a Kolmogorov-like scaling in the inertial range, roughly located between $10^{-4}$ Hz and $10^{-1}$ Hz.
The average spectral index is $\alpha_{\rm{MHD}}=1.68\pm0.05$, the error being the standard deviation over the $48$ cases, and is in agreement with the literature~\citep{brunocarbone2013}. Figure~\ref{fig:f6} shows the histogram of the exponents (in red), revealing the narrow dispersion associated with the magnetohydrodynamic inertial range spectral decay.

At lower frequencies, the survey provided the following possible  behaviours:
(i) the Kolmogorov inertial range extends to all observed frequencies (observed in 4 samples, or 8\% of the cases; see the example in Fig.~\ref{fig:f4});
(ii) there is evidence of a low-frequency spectral break, but no well-defined large-scale power law (6 samples, 13\%);
(iii) there is a well-defined power law for at least one frequency decade below the low-frequency break (38 samples, 79\%; see the example in Fig.~\ref{fig:f2}); and
(iv) the spectrum shows a flat (white noise) region at the lower frequencies, after a low-frequency break---thus excluding the cases of group  (i)---but irrespective of the presence or absence of power-law scaling---thus joining the cases of groups (ii) and (iii)---(12 samples, 25\%).

%\LEt{ see note 4 }
When a power law  is observed, the average spectral index is $\alpha_{\rm{low}}=1.13\pm0.1$, and its distribution given in panel b of Fig.~\ref{fig:f6} shows a broader variability than for the inertial range.
The predominant number of cases with a power law observed above (i.e. group (ii), including nearly $80\%$ of the samples) demonstrates that extended intervals of slow solar wind with low compressibility are predominantly characterized by a typical low-frequency $1/f$ spectral range.

\section{Discussion}

The reasons for different spectral behaviour observed in the slow wind intervals analysed for this work could be due to the different characteristics of the fluctuations forming the turbulence spectrum.
To explore this possibility, for each of the $48$ slow wind streams we evaluated the level of compressibility and Alfv\'{e}nicity of the turbulent fluctuations, but the results were similar to those shown in Figure~\ref{fig:f5}, not showing any particular correlation between these parameters and the spectral form.

\begin{figure*}
        \includegraphics[width=15cm]{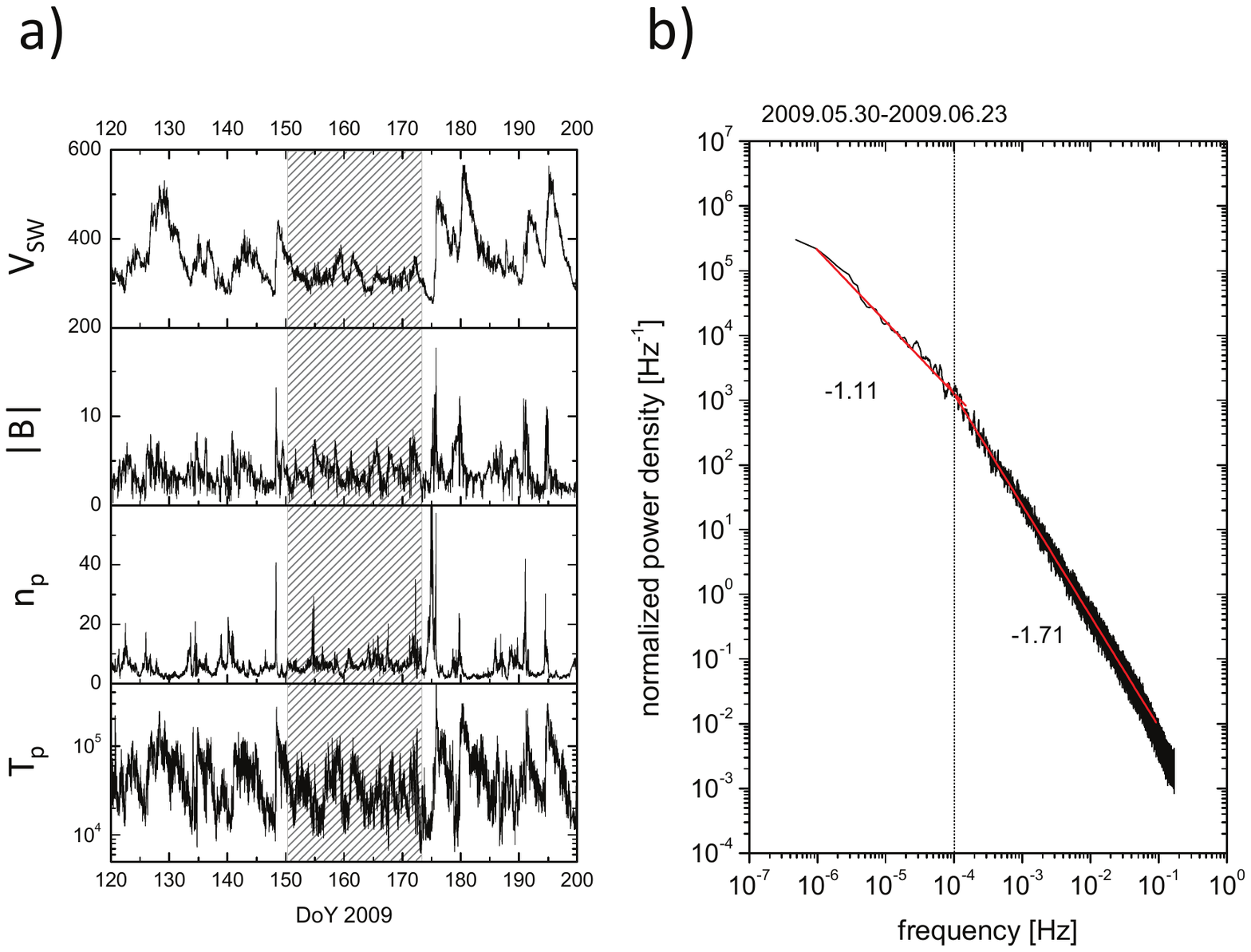}
    \caption{Panel a: One-minute averages of solar wind parameters: wind speed [km s$^{-1}$], magnetic field intensity [nT], proton number density [cm$^{-3}$], and proton temperature [K] are shown (from top to bottom). Panel b: Trace of the power density spectral matrix of magnetic field fluctuations, normalized to the square value of the mean magnetic field intensity, relative to the time interval highlighted by the shaded area in the left panel.}
\label{fig:f2}
\end{figure*}

\begin{figure*}
\includegraphics[width=15cm]{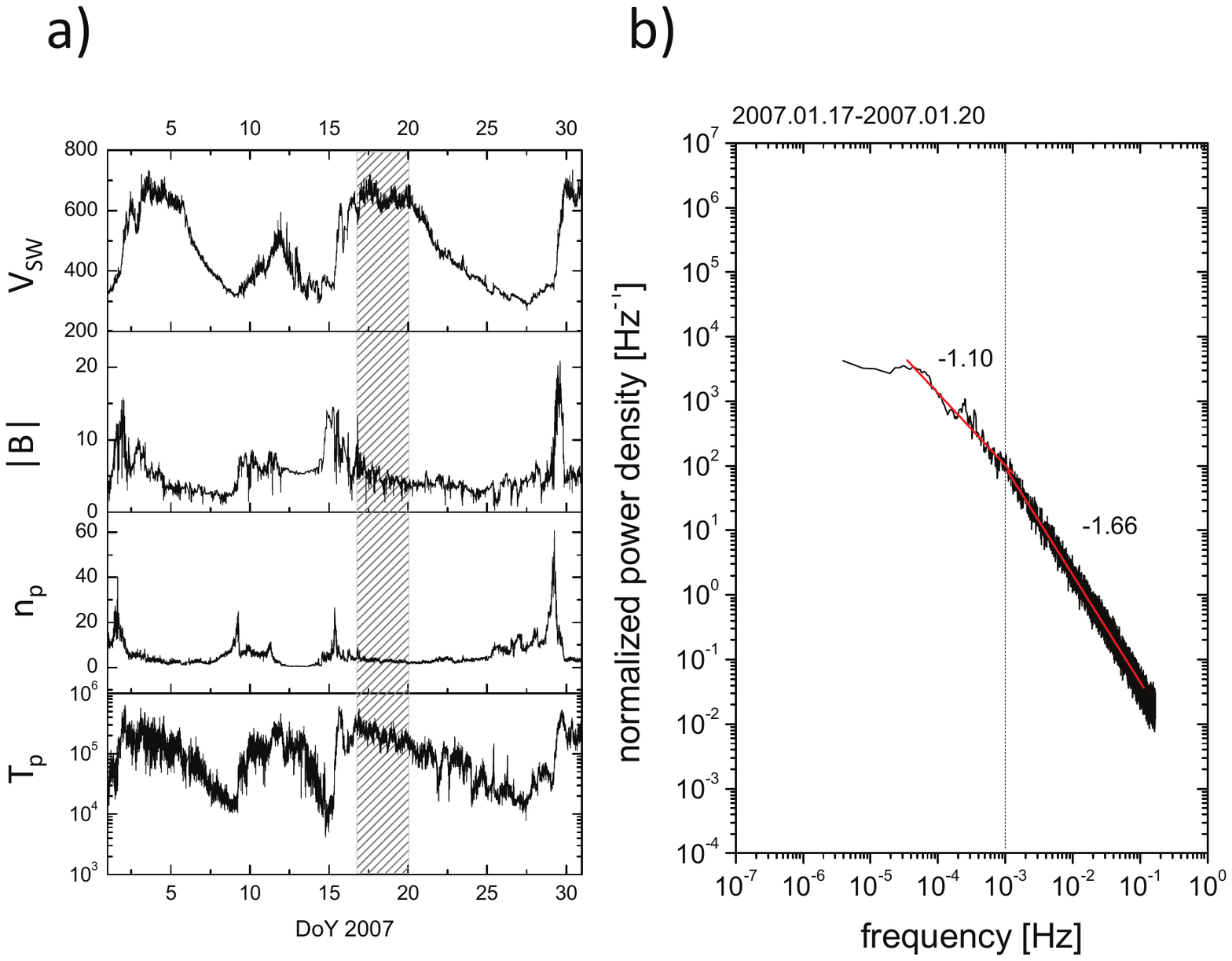}
\caption{Panel a: One-minute averages of solar wind parameters for a typical fast wind interval, in the same format as in Fig. \ref{fig:f2}. Panel b: Trace of the power density spectral matrix of magnetic field fluctuations, normalized to the square value of the mean magnetic field intensity, in the same format as in Fig. \ref{fig:f2}.} \label{fig:f3}
\end{figure*}

\begin{figure*}
\includegraphics[width=15cm]{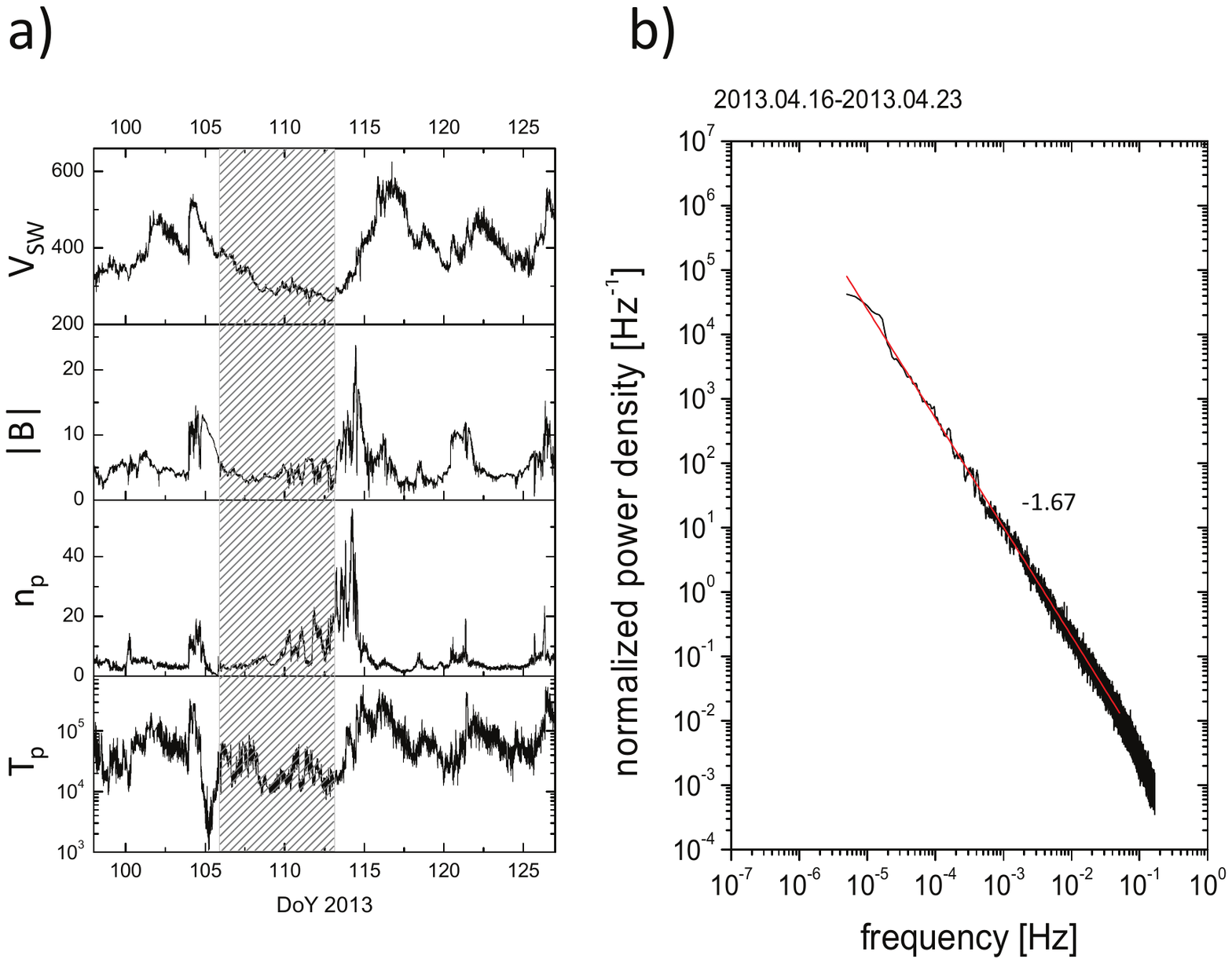}
\caption{Panel a: One-minute averages of solar wind parameters, for one of the 48 selected slow wind intervals, in the same format as in Fig. \ref{fig:f2}. Panel b: Trace of the power density spectral matrix of magnetic field fluctuations, normalized to the square value of the mean magnetic field intensity, in the same format as in Fig. \ref{fig:f2}.} \label{fig:f4}
\end{figure*}

\begin{figure}
\includegraphics[width=10cm]{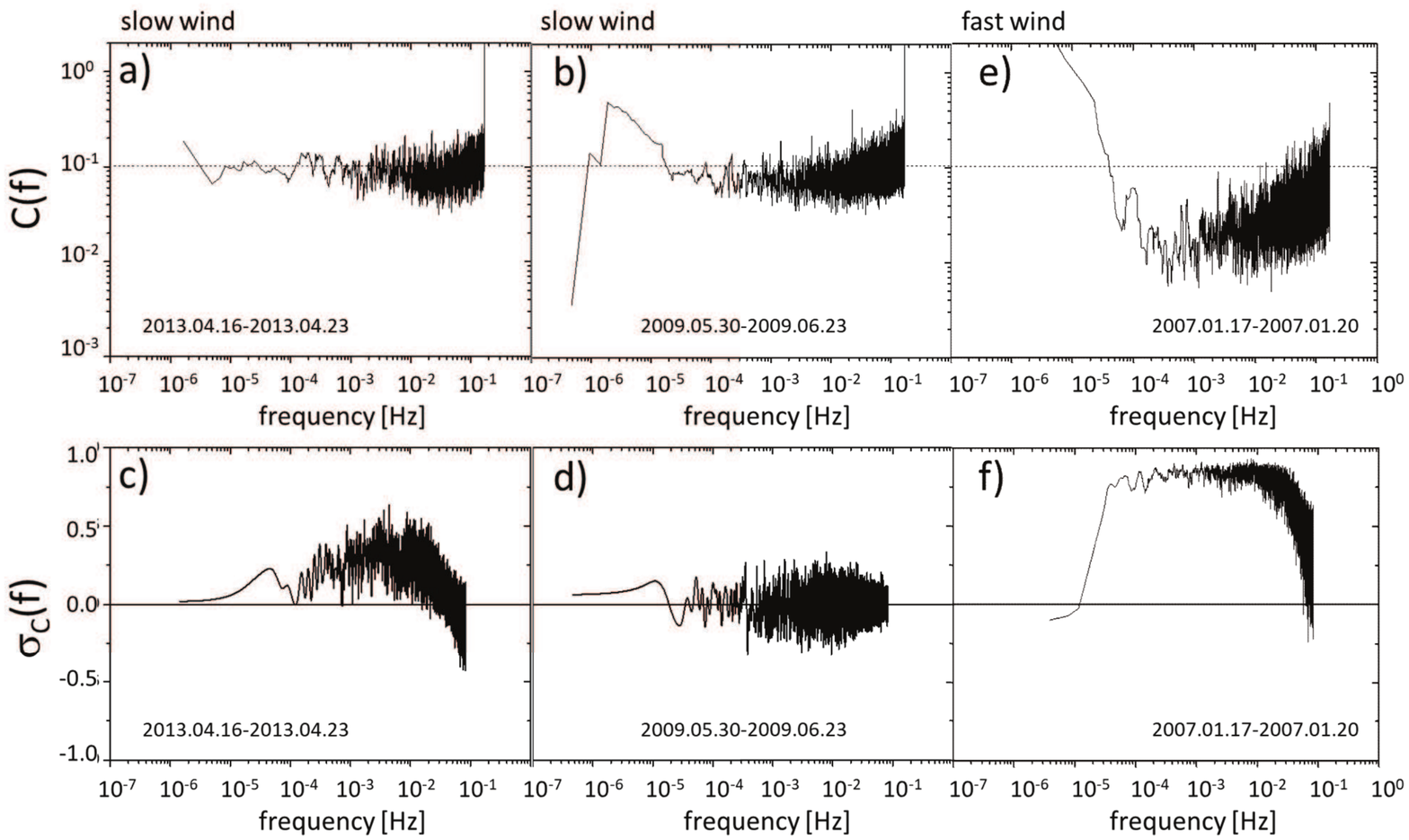}
\caption{Magnetic compressibility and Alfv\'{e}nicity for the slow wind intervals shown in Figs. \ref{fig:f4} and \ref{fig:f2}, and for the fast wind interval shown in Fig. \ref{fig:f3}, as also indicated by the time interval reported in each panel.}\label{fig:f5}
\end{figure}

\begin{figure*}
\includegraphics[width=10cm]{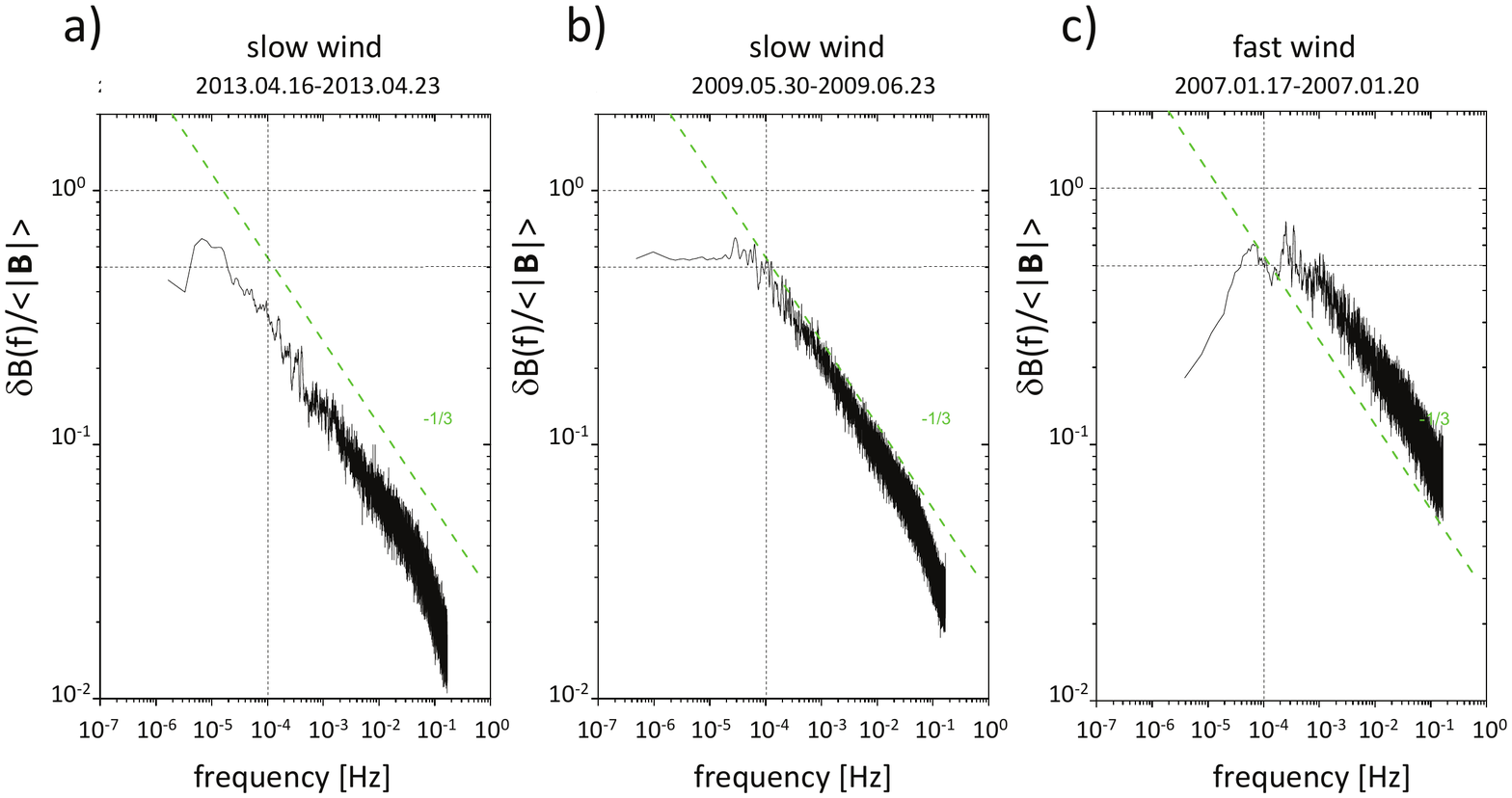}
\caption{From left to right: Normalized amplitude of magnetic field fluctuations for the slow wind time intervals shown in Fig. \ref{fig:f4} and Fig. \ref{fig:f2}, and for the fast wind interval shown in  Fig. \ref{fig:f3}.}\label{fig:f7}
\end{figure*}

At this point, we checked whether a saturation effect of the fluctuations, similar to that suggested by \cite{matteini2018} for the fast Alfv\'{e}nic wind, could also play a role  in the slow wind spectral break.
Thus, in order to explore this possibility, differently from \cite{matteini2018} who used first-order structure functions, we estimated the amplitude of each Fourier mode using the simple relation that binds together the power of a given fluctuation and the amplitude of the fluctuation $\delta B(f)$, by means of the Fourier power spectral density $S(f)$, namely $\delta B(f)=\sqrt{2fS(f)}$. These values were successively normalized to the corresponding local magnetic field average within each interval $\langle |\mathbf{B}|\rangle,$ and are shown in Fig.~\ref{fig:f7}. Panel a refers to the slow wind interval of Fig.~\ref{fig:f4}, characterized by an extended Kolmogorov spectrum and no $1/f$ range; panel b refers to the slow wind interval of Fig.~\ref{fig:f2}, which shows the $1/f$ scaling; and panel c refers to the fast wind interval of Fig.~\ref{fig:f3}, which shows (as expected for the fast wind) an extended Kolmogorov spectrum and a clear $1/f$ scaling.
In all panels, the green dashed line is an arbitrary reference level with the expected $f^{-1/3}$ scaling for the amplitude $\delta B(f)$ of the Fourier modes.
This line has been drawn to facilitate the comparison between the different levels of $\delta B(f)/\langle|\mathbf{B}|\rangle$ in the three panels.

Since these plots directly derive from their corresponding normalized power density spectra shown in Figs.~\ref{fig:f2},~\ref{fig:f3}, and~\ref{fig:f4} we observe a flattening in the same frequency range where the power density spectra shows the $1/f$ scaling. In other words, the flattening indicates that the amplitude of the Fourier modes has reached a limit, i.e., below a certain frequency the fluctuations are saturated. This particular condition is reached at higher and higher frequencies depending on the relative amplitude of the fluctuations with respect to the local field. This is evident moving from panel a to panel c of Fig.~\ref{fig:f7}.

As already recalled before, \cite{matteini2018} suggested that the $1/f$ scaling observed in fast wind magnetic field might be the consequence of the large-scale saturation of the fluctuations. In this perspective, the amplitude of the fluctuations would be limited by the magnitude of the local magnetic field.
The results shown here would expand this interpretation of the $1/f$ range to the slow solar wind.

\begin{figure*}
\includegraphics[width=11cm]{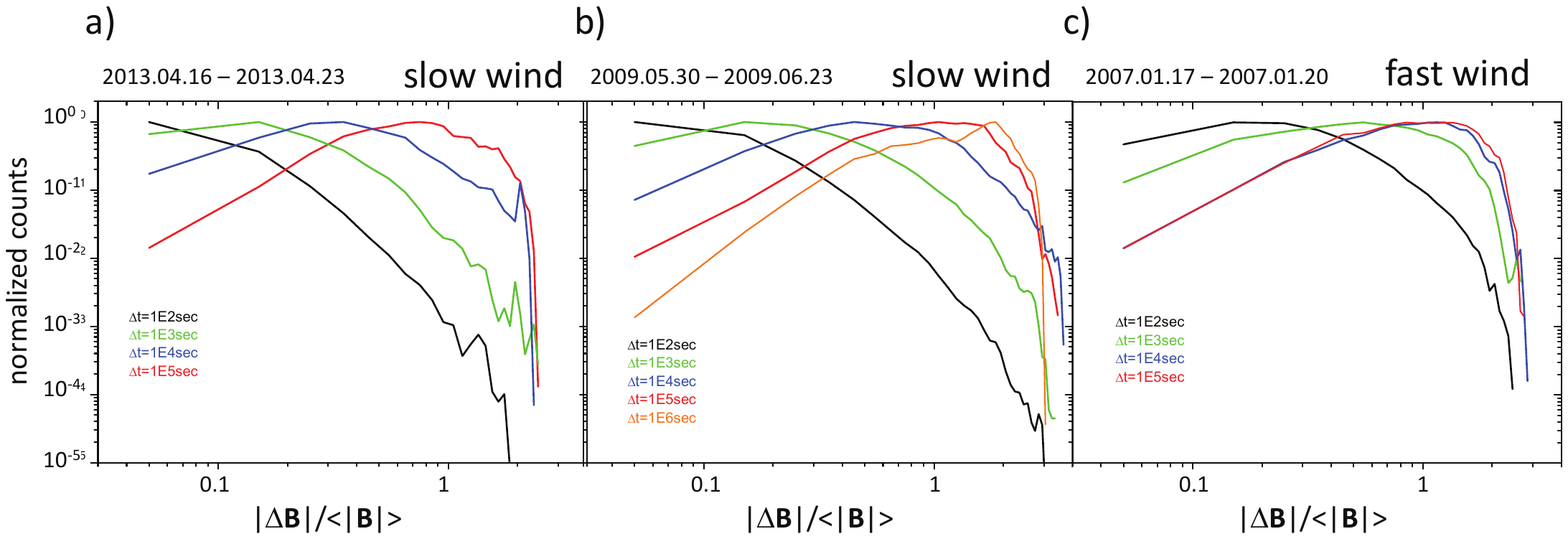}
\caption{Histograms of the amplitude of the magnetic field fluctuations normalized to the average value of the field intensity within the selected time interval. Each curve corresponds to a different timescale, as indicated by the color-coding, and is normalized to its maximum value.}\label{fig:f8}
\end{figure*}

In order to strengthen this interpretation with additional experimental evidence we show, in Fig.~\ref{fig:f8}, the histograms of the normalized amplitude of the magnetic field fluctuations $|\mathbf{B}(t+\Delta t)-\mathbf{B}(t)|/\langle|\mathbf{B}|\rangle$, where $\Delta t$ is the timescale and $\langle|\mathbf{B}|\rangle$ is the average value of the field intensity within the selected time interval. If the magnetic field intensity did not change with time, $|\mathbf{B}(t+\Delta t)-\mathbf{B}(t)|/\langle|\mathbf{B}|\rangle$ would have a limiting value of $2$.
Each curve in each panel has been normalized to its maximum value. The three panels a, b, and c correspond to the three different time intervals described in Fig.~\ref{fig:f4}, Fig.~\ref{fig:f2}, and Fig.~\ref{fig:f3}, respectively. The different timescales are indicated by the colour-coding shown in each panel. Panel a corresponds to the slow wind interval without magnetic field spectral break,  while panel b corresponds to the time interval which shows the break. The different histograms in panels a, b, and c refer to four different timescales, namely $10^2, 10^3, 10^4$, and  $10^5$ sec. In addition, given the remarkable length of the corresponding time interval, panel b also shows  the histogram for the timescale $10^6$ sec. It is interesting to note that  increasing the timescale moves the peak of the corresponding histogram  to higher values of $|\Delta \mathbf{B}|/\langle|\mathbf{B}|\rangle$ in each panel. However, only for panels b and c do the curves display their maximum value around  $|\Delta \mathbf{B}|/\langle|\mathbf{B}|\rangle \sim 2$ and, in particular, only the histograms corresponding to the timescales falling within the $1/f$ spectral range in Fig.~\ref{fig:f2}b (slow wind) and  Fig.~\ref{fig:f3}b (fast wind) tend to collapse on each other. This phenomenon is a clear indication that fluctuations become saturated starting at the timescale where the peak of the corresponding histogram is around $2$. Obviously, this limiting value can be larger than $2$ depending on the compressive level of the fluctuations, being exactly $2$ only if the magnetic field vector fluctuates on the surface of a sphere of constant radius.
To this regard, it is worth  mentioning that \citet{{tsurutani2018}}, studying a low-compression high-speed stream, did find Alfv\'{e}nic fluctuations whose amplitudes were equal to the entire magnetic field strength.
However, we like to remark that for the slow wind cases analysed in this work, the saturation effect is not due to Alfv\'{e}nic fluctuations but rather to different local orientations of the magnetic field reflecting the background field within adjacent static structures advected by the wind \citep{bruno2004}.
These results show that the phenomenon of saturation of magnetic field fluctuations is common to both fast and slow winds. In other words, it appears  that it is not the nature of the  fluctuations $|\delta \mathbf{B}|$ but their amplitude relative to the background field intensity $|\mathbf{B}|$  that causes saturation and, consequently, the formation of the 1/f spectral range.

However, there are still important differences between fast and slow solar wind turbulence, which highlight the different natures of the turbulence fluctuations.
The first difference is that the break in the fast wind is located at higher frequencies, since magnetic field fluctuations are Alfv\'{e}nic in nature and, as such, are much larger than magnetic field fluctuations in slow wind  \citep{brunocarbone2013}.

\begin{figure*}
\includegraphics[width=11cm]{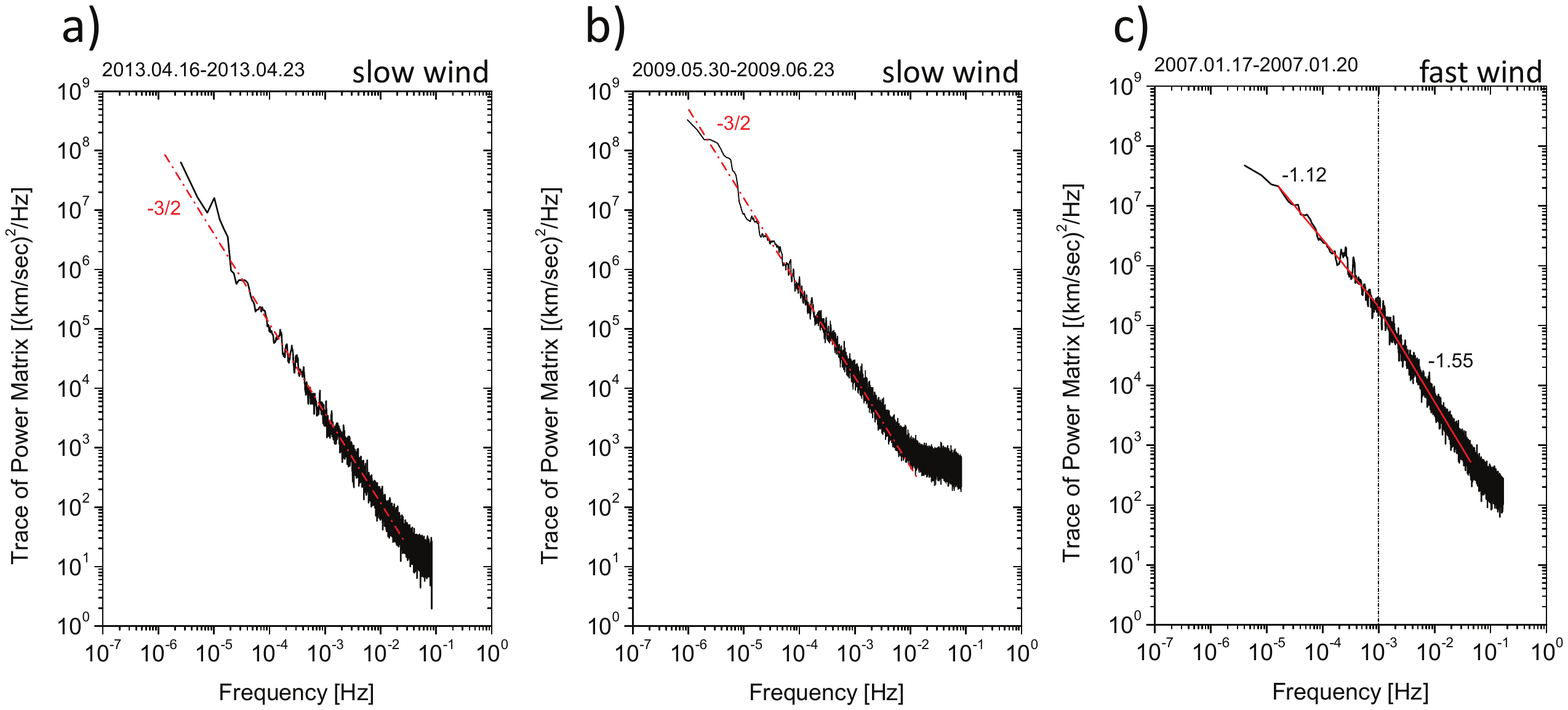}
\caption{From left to right: Trace of the power density spectral matrix of velocity fluctuations for the slow wind time intervals shown in Fig. \ref{fig:f4} and Fig. \ref{fig:f2}, and for the fast wind interval shown in  Fig. \ref{fig:f3}.}\label{fig:f9}

\end{figure*}

Another relevant difference can be noticed in the behaviour of the velocity fluctuations. Figure~\ref{fig:f9} shows the three velocity spectra corresponding to the same time intervals discussed above. It is evident that there is no $1/f$ range in either of the two slow wind samples (panels a and b). Velocity and magnetic field fluctuations are decoupled, as expected for turbulence characterized by low or absent Alfv\'enicity.
On the other hand, as shown in panel c of the same figure, the velocity spectrum for the Alfv\'{e}nic fast solar wind typically displays the same properties as the magnetic field, including the $1/f$ low-frequency spectrum, since the two fields are strongly correlated. In this case, the saturation controls the magnetic fluctuations, and the velocity fluctuations would adapt to maintain the Alfv\'enic correlation.

The idea that solar wind fluctuations at hourly scale, i.e. within the $1/f$ scaling range that we observe in the inner heliosphere, might be saturated dates back to early {in situ} observations by \citet{belcher1974}, \citet{mariani1978,mariani1979} and \citet{villante1980}.
%\LEt{ To integrate your references properly into the main text, please separate them with commas  and set off the last reference with "and". This is a LaTeX command error (citet and citep) that I cannot fix for you with the program I work with. }
In fact, it was found that the ratio between the total variance of the fluctuations $\sigma^2 =\sum_{i=x,y,z}{\sigma_i}^2$ and the square value of the local magnetic field intensity $|\mathbf{B}|^2$, within fast wind, was essentially independent of the heliocentric distance. This evidence was interpreted in terms of fluctuations for which the ratio of their energy density  to that of the background magnetic field would saturate to some constant value. Actually, while these authors found that $\sigma^2$
decreases as $\sim R^{-3}$, \citet{behannon1978} showed that the radial dependence of the interplanetary magnetic field magnitude is nicely approximated by $\sim R^{-1.5}$.

This last value is remarkably close to the radial dependence of the low-frequency break found by \citet{brunocarbone2013} for the fast wind and suggests a robust link between the radial dependence of the field intensity $|\mathbf{B}|$ and the possible saturation effect discussed above.

%%\subsection{Animations}

In order to highlight the link existing between the presence of a low-frequency break and the saturation of the amplitude of the fluctuations, we selected $14$ time intervals out of the original $48$ and made an animation which shows the progressive appearance of the low-frequency break as the relative fluctuations increase in amplitude
(\verb"movie.mp4"). From the animation, the considerable variability in the frequency break location can also be noticed, which, as already noted, does not obey the estimates provided by the radial dependence $R^{-1.52}$ found by  \citet{brunocarbone2013} for the fast wind.

\begin{acknowledgements}
The authors acknowledge stimulating discussions with L. Matteini, S. Landi, A. Verdini and M. Velli.
We especially thank B. Tsurutani for giving helpful comments on a version of the paper.
This work was partially supported by the Italian Space Agency (ASI) under contract ACCORDO ATTUATIVO n. 2018-30-HH.O
The Wind
%\sout{WIND}
data used in this work are publicly available from the NASA-CDAWeb repository (\verb"https://cdaweb.sci.gsfc.nasa.gov").
Results from the data analysis presented in this paper are directly available from the authors.
\end{acknowledgements}

\begin{thebibliography}{}


    \bibitem[Behannon(1978)]{behannon1978}
Behannon, K.~W. 1978, Rev. Geophys. and Space Phys., 16, 125-145.
        \bibitem[Bavassano et al., 1982]{bavassano1982}
Bavassano, B., Dobrowolny, M., Mariani, F., et al. 1982, \solphys, 78, 373
    \bibitem[Belcher \& Burchsted(1974)]{belcher1974}
Belcher, J.~W., \& Burchsted, R. 1974, \jgr, 79, 4765
    \bibitem[Bruno et al.(2004)]{bruno2004}
Bruno, R., Carbone, V., Primavera, L., Malara, F., Sorriso-Valvo, L., Bavassano, B., Veltri, P.\ 2004.\ Ann. Geophys. 22, 3751-3769.
    \bibitem[Bruno et al.(2009)]{bruno2009}
Bruno, R., Carbone, V., V\"or\"os, Z., et al. 2009, Earth Moon and Planets, 104, 101
        \bibitem[Bruno \& Carbone, 2013]{brunocarbone2013}
Bruno, R., \& Carbone, V. 2013, Living Rev. Solar Phys. 10, 2
        \bibitem[Bruno \& Trenchi, 2014]{bruno2014a}
Bruno, R., \& Trenchi, L. 2014, \apjl, 787, L24
    \bibitem[Consolini et al.(2015)]{consolini2015}
Consolini, G., De Marco, R., \& Carbone, V. 2015, \apj, 809, 21
    \bibitem[Costa et al., 2014]{Costa}
Costa A., Osborne, A.~R., Resio, D.~T., et al. 2014, \prl, 113(10):108501
    \bibitem[D'Amicis et al.(2019)]{raffaella2019}
D'Amicis, R., Matteini, L., \& Bruno, R. 2019, \mnras, 483, 4665
    \bibitem[D'Amicis \& Bruno(2015)]{raffaella2015}
D'Amicis, R., \& Bruno, R. 2015, \apj, 805, 84
    \bibitem[Dmitruk et al., 2002]{dmitruk2002}
Dmitruk, P., Matthaeus, W.~H., Milano, L.~J., et al. 2002, \apj, 575, 571
        \bibitem[Dmitruk et al., 2004]{dmitruk2004}
Dmitruk, P., Matthaeus, W.~H., \& Seenu, N. 2004, \apj, 617, 667
    \bibitem[Dmitruk \& Matthaeus, 2007]{dmitruk2007}
Dmitruk, P., \& Matthaeus, W.~H. 2007, \pre, 76, 036305
    \bibitem[Dmitruk et al., 2011]{dmitruk2011}
Dmitruk, P., Mininni, P.~D., Pouquet, A., et al. 2011, \pre, 83, 066318
    \bibitem[Fraedrich \& Blender, 2003]{Fraedrich}
Fraedrich, K., \& Blender, R. 2003, \prl, 90, 108501
    \bibitem[Frisch, 1995]{frisch}
Frisch, U. 1995, Turbulence.~The legacy of A.~N.~Kolmogorov., by Frisch, U.,~ Cambridge University Press, Cambridge (UK), 1995, XIII + 296 p., ISBN 0-521-45103-5
    \bibitem[Herault et al., 2015]{Herault}
Herault J., P\'etr\'elis, F., \& Fauve, S. 2015, Europhys. Lett. 111 (4), 44002
    \bibitem[Horbury et al.(1996)]{horbury1996}
Horbury, T.~S., Balogh, A., Forsyth, R.~J., \& Smith, E.~J. 1996, \aap, 316, 333
    \bibitem[Kolmogorov, 1941]{K41}
Kolmogorov, A.~N. 1941, Dokl. Akad. Nauk. SSSR 30, 301
        \bibitem[Lepping et al., 1995]{lepping1995}
Lepping, R.~P., Acu\~{n}a, M.~H., Burlaga, L.~F., et al., 1995, \ssr, 71, 207
        \bibitem[Lin et al., 1995]{lin1995}
Lin, R.~P., Anderson, K.~A., Ashford, S., et al., 1995, \ssr, 71, 125
    \bibitem[Mariani et al.(1979)]{mariani1979}
Mariani, F., Villante, U., Bruno, R., et al. 1979, \solphys, 63, 411
    \bibitem[Mariani et al.(1978)]{mariani1978}
Mariani, F., Ness, N.~F., Burlaga, L.~F., et al. 1978, \jgr, 83, 5161
    \bibitem[Matteini et al., 2018]{matteini2018}
Matteini, L., Stansby, D., Horbury, T.~S., \& Chen, C.~H.~K. 2018, \apj, 869, L32
    \bibitem[Matthaeus \& Goldstein(1986)]{matthaeus1986}
Matthaeus, W.~H., \& Goldstein, M.~L. 1986, \prl, 57, 495
        \bibitem[Matthaeus et al., 2005]{matthaeus2005}
Matthaeus, W.~H., Dasso, S, Weygand, J.~M., et al. 2005, \prl, 95, 231101
    \bibitem[Matthaeus et al.(2007)]{matthaeus2007}
Matthaeus, W.~H., Breech, B., Dmitruk, P., et al. 2007, \apj, 657, L121
    \bibitem[Montroll \& Shlesinger(1982)]{montroll1982}
Montroll, E.~W., \& Shlesinger, M.~F. 1982, Proc. National Academy of Science 79, 3380
    \bibitem[Nakagawa \& Levine(1974)]{nakagawa1974}
Nakagawa, Y., \& Levine, R.~H. 1974, \apj, 190, 441
    \bibitem[Pereira et al., 2019]{Pereira}
Pereira, M., Gissinger, C., \& Fauve, S. 2019, \pre, 99, 023106
    \bibitem[Telloni et al.(2015)]{telloni2015}
Telloni, D., Bruno, R., \& Trenchi, L. 2015, \apj, 805, 46
    \bibitem[Tenerani \& Velli(2017)]{tenerani2017}
Tenerani, A., \& Velli, M. 2017, \apj, 843, 26
    \bibitem[Tsurutani et al.(2018)]{tsurutani2018}
Tsurutani, B.~T., Lakhina, G.~S., Sen, A., et al. 2018.\jgr 123, 2458
        \bibitem[Tu \& Marsch, 1995]{tu1995}
Tu, C. Y., \& Marsch, E. 1995, \ssr, 73, 1
    \bibitem[Velli et al.(1989)]{velli1989}
Velli, M., Grappin, R., \& Mangeney, A. 1989, \prl, 63, 1807
    \bibitem[Verdini et al.(2012)]{verdini2012}
Verdini, A., Grappin, R., Pinto, R., \& Velli, M. 2012, \apj, 750, L33.
    \bibitem[Villante(1980)]{villante1980}
Villante, U. 1980, \jgr, 85, 6869

\end{thebibliography}
\end{document}